# Optical force-induced nonlinearity and self-guiding of light in human red blood cell suspensions

Rekha Gautam[1,2*], Yinxiao Xiang[1,3*], Josh Lamstein[1*], Yi Liang[1,4], Anna Bezryadina[1,5], Guo Liang[1], Tobias Hansson[6,7], Benjamin Wetzel[6,8], Daryl Preece[9], Adam White[1], Matthew Silverman[10], Susan Kazarian[10], Jingjun Xu[3], Roberto Morandotti[6,11,12], and Zhigang Chen[1,3]

[1]*Department of Physics and Astronomy, San Francisco State University, San Francisco, CA 94132,USA*
[2]*Department of Biomedical Engineering, Vanderbilt University, Nashville, TN 37240,USA*
[3]*MOE Key Lab of Weak-Light Nonlinear Photonics, TEDA Applied Physics Institute and School of Physics, Nankai University, Tianjin 300457, China*
[4]*Guangxi Key Lab for Relativistic Astrophysics, Guangxi Colleges and Universities Key Lab of Novel Energy Materials and Related Technology, School of Physical Science and Technology, Guangxi University, Nanning, Guangxi 530004, China*
[5]*Department of Physics and Astronomy, California State University Northridge, Northridge, CA 91330, USA*
[6]*Institut National de la Recherche Scientifique, Université du Québec, Varennes, Québec J3X 1S2, Canada*
[7]*Department of Physics, Chemistry and Biology, Linköping University, Linköping SE-581 83, Sweden*
[8]*School of Mathematical and Physical Sciences, University of Sussex, Sussex House, Falmer, Brighton BN1 9RH, UK*
[9]*Department of Biomedical Engineering, University of California Irvine, Irvine, CA,USA*
[10]*Clinical Laboratory Science Program, San Francisco State University, San Francisco, CA 94132, USA*
[11]*Institute of Fundamental and Frontier Sciences, University of Electronic Science and Tech. of China, Chengdu 610054, China*
[12]*ITMO University, Saint Petersburg 197101, Russia*
*These authors made equal contribution.
Corresponding author: *zhigang@sfsu.edu, yinxiaoermao@nankai.edu.cn*

**Abstract:**

Osmotic conditions play an important role in the cell properties of human red blood cells (RBCs), which are crucial for the pathological analysis of some blood diseases such as malaria. Over the past decades, numerous efforts have mainly focused on the study of the RBC biomechanical properties that arise from the unique deformability of erythrocytes. Here, we demonstrate nonlinear optical effects from human RBCs suspended in different osmotic solutions. Specifically, we observe self-trapping and scattering-resistant nonlinear propagation of a laser beam through RBC suspensions under all three osmotic conditions, where the strength of the optical nonlinearity increases with osmotic pressure on the cells. This tunable nonlinearity is attributed to optical forces, particularly the forward scattering and gradient forces. Interestingly, in aged blood samples (with lysed cells), a notably different nonlinear behavior is observed due to the presence of free hemoglobin. We use a theoretical model with an optical force-mediated nonlocal nonlinearity to explain the experimental observations. Our work on light self-guiding through scattering bio-soft-matter may introduce new photonic tools for noninvasive biomedical imaging and medical diagnosis.

**Introduction**

When a light beam enters a turbid medium such as blood or other biological fluids, it experiences multiple scattering and loses its power and original directionality. This issue has hampered many applications where transmission, focusing, or imaging through scattering media is desirable, which also motivated a great deal of interest in developing new methods and techniques to overcome these hurdles [1-8]. In particular, for medical and biological applications, low-loss transmission of light is a requisite for deep-tissue imaging, localized laser treatments, and internal optical control of particles or microrobots for medical diagnosis and treatment, to name a few [9-15]. Until now, significant efforts have been made to mainly enhance light transmission by understanding light-matter interactions in the linear optical domain [16,17], but it remains largely unexplored in the nonlinear domain. In fact, it is commonly thought that light in biological environments only exhibits a negligible or weak optical nonlinearity, so the required power must be very high to change the refractive index of a biological medium with a laser beam, which can cause photodamage. In recent years, typical nonlinear effects have been induced by ultrafast laser pulses such as multiphoton excitation, second and third harmonic generation, and Kerr effects to obtain better resolution and deeper penetration through scattering bio-media [18-20]. Quite recently, we have achieved low-loss propagation of light through bacterial suspensions by exploiting the nonlinear optical properties of marine bacteria while demonstrating that the viability of the cells remains intact [21]. This achievement clearly indicates that even with a continuous-wave (CW) laser operating at relative low intensities, biological media can exhibit appreciable optical nonlinearity.

Human red blood cells (RBCs) in normal conditions are disc-shaped malleable cells, approximately 8 μm in diameter and 2 μm in thickness, which have a spatially uniform refractive index because they lack nuclei and most organelles [22-24]. To enable the passage through veins and narrow microcapillaries, RBCs exhibit distinctive deformability following the application of an external force. Deformation can also be elicited by modifying the liquid buffer osmolarity. As an exemplary application of such a unique feature, it has been demonstrated that RBCs can be used as tunable optofluidic microlenses [22]. Important to both *in vitro* and *in vivo* disease diagnostics, the optical properties of RBCs depend on the shape and refractive index of cells. The RBC refractive index is mainly determined by hemoglobin (Hb), which is the largest part of the erythrocyte dry content by weight [25]. The refractive index increases if the cell volume decreases with varying osmotic conditions [24,26-30]. The physical properties of RBCs involving cell size and shape are often closely related to pathophysiological conditions such as sickle cell anemia, malaria, and sepsis [31-35]. Considering the intrinsic fundamental interest, the ability of RBCs to react to changes in different osmotic environments make them ideal candidates for the study of light scattering by varying the refractive index and shape of the cells [29,32,33].

In this work, we study the optical nonlinearity of RBCs in dynamic, monodisperse, colloidal suspensions under different osmotic conditions, and we demonstrate the nonlinear self-trapping of light over centimeter propagation distances through scattering RBC suspensions. If there is no nonlinearity in such suspensions, a passing laser beam should linearly diffract regardless of the optical power. However, increasing the laser beam power to only a few hundreds of mW (as opposed to 3 W in colloidal suspensions of marine bacteria featured by a smaller size [21,36]), we observe that the beam dramatically self-focuses in all three osmotic conditions and forms a self-trapped channel, which is similar to an optical spatial soliton [37]. Interestingly, the optical nonlinearity is also tunable via osmosis, and we find that in fresh blood samples, the nonlinearity increases with the osmotic pressure outside the cells. Intuitively, one can understand such nonlinear beam dynamics from the optical forces that act on the RBCs: the gradient force attracts the RBCs towards the beam center, while the scattering force pushes them forward. When there are sufficiently many cells along the beam path, an effective waveguide is formed as RBCs have a higher index of refraction than the background buffer solution. The theoretical analysis of this behavior is not trivial because the cells do not have a fixed shape, size, or refractive index. In addition, the situation can be further complicated due to the lysed RBCs in aged samples [38], where free Hb plays an active role in the value of the optical nonlinearity along with strong thermal effects. Nevertheless, a theoretical model for the optical force-mediated nonlocal nonlinearity is proposed to analyze the beam dynamics, where the results are qualitatively consistent with our experimental observations. The enhanced transmission of light through scattering blood cells driven by optical forces may find applications in medical diagnostics. For example, the changes in optical forces with the cell density and morphology can provide a powerful noninvasive tool to sort different cells according to the stage of a given disease [27].

**Results**

The typical experimental results for the nonlinear beam propagation in RBC suspensions are presented in Fig. 1, where the plots in (a-c) illustrate the propagation of a laser beam through an RBC suspension in different osmotic conditions. The focused laser beam ($\lambda$=532 nm) normally diffracts in the phosphate saline (PBS) background solution alone (i.e., without RBCs) and exhibits no nonlinear self-action at any tested laser power of 10-700 mW. However, when the laser power increases, a nonlinear optical response is observed in the RBC suspensions for all three osmotic conditions (isotonic, hypotonic and hypertonic), which causes the self-trapping of the light beam and the formation of an optically induced biological waveguide [21]. A side-view propagation of such a self-trapped optical beam is shown in Fig. 1(d), and an animation of the dynamical process is shown in the Supplementary Material. Careful measurements show that the RBCs suspended in different osmotic solutions exhibit different strengths in terms of optical nonlinearities, where a laser beam with the same input size requires different laser powers to achieve self-trapping. As shown in Figs. 1(e-h), the beam first normally diffracts at a low power of 10 mW and experiences strong scattering due to a random distribution of nonspherically shaped RBCs. In the isotonic solution (n~1.42 for RBCs) [22,42], the optimal self-trapping of light (i.e., focusing to the smallest possible spot size) is observed when the beam power reaches approximately 300 mW (Fig. 1i). Interestingly, nonlinear self-trapping occurs at a slightly higher power of 350 mW in the hypotonic solution (n~1.38) [22,42] (Fig. 1j), and instead it happens at a slightly lower power of 200 mW (Figs. 1k-1l) in the hypertonic solution (n~1.44) [22,42]. These results clearly show that RBCs have an appreciable nonlinear response, and the optical nonlinearity is tunable because it can change when the optical forces vary under different osmotic conditions. We emphasize that it is not simply the result of a better laser transmission through scattering media at higher powers, since both optical gradient and scattering forces play an active role in the self-trapping, as elaborated below.

To show that the observed phenomenon arises from nonlinear self-focusing, we measure the normalized transmission (output/input power) as a function of the input beam power. These results are summarized in Fig. 2a, where for direct comparison, we intentionally set the initial transmission to be 1.00 for all conditions; the linear propagation in the PBS-only solution is the control data. Clearly, different nonlinear upward trends of the normalized transmission are observed for different osmotic conditions, which indicates a dissimilar strength of optical nonlinearity and self-focusing of the optical beam (as shown in Fig. 2a). The initial transmission varies for the RBCs under three different osmotic conditions because of the difference in absorption and shape-dependent scattering [22], as detailed in the Supplementary Material. Figure 2b illustrates the changes in output beam size due to nonlinear self-action through different osmotic solutions of fresh blood samples (as a function of the beam power). The beam size remains fairly constant in the background medium without RBCs when the power increases, which exhibits no appreciable self-focusing. However, in all three RBC suspensions, when the power increases, the beam size first dramatically decreases due to nonlinear self-trapping and subsequently reaches a minimum at a few hundred mW before starting to increase again due to thermal effects [21,43,44]. Specifically, in the hypotonic solution, the RBCs are in a "swollen" state, so their effective refractive index decreases when the water-to-Hb ratio increases. In this case, the power required to focus the beam to its smallest spot size (approximately 350 mW) is higher than those for the other two cases, and the power transmission is approximately 28% after propagating approximately 3 cm through the RBC suspension. In contrast, in the hypertonic solution, where the RBCs are in a "shrunk" state, the cells become denser, and the effective refractive index increases due to a reduced water-to-Hb ratio. In this case, the beam has a lower threshold power for self-trapping (approximately 200 mW), since both the optical gradient and scattering forces are higher than the other two cases. The power transmission is only approximately 20% since the suspension becomes comparatively more turbid, although the number of cells in the suspension is approximately identical. As expected, in the "normal" state of the isotonic condition, the RBCs show an intermediate behavior with respect to the self-trapping threshold power and normalized transmission. Surprisingly, the experiments performed using the same blood samples after the cells had been stored for two weeks exhibited notably different outcomes (Fig. 2c), particularly for the hypotonic suspensions as we discuss below. It is important to note that for the fresh sample, the health of the RBCs was assessed after the laser illumination, and no significant photodamage was observed for the laser power in our experiments, as discussed in the Supplementary Material.

To further corroborate the above argument about the difference in optical force-mediated nonlinearity in RBC suspensions under varying osmotic conditions, we used optical tweezers to directly observe the cell motion under a microscope (NA=1.3) and to do direct force measurement of single cells in the trap. The typical experimental results are presented in Fig. 3 with a 960-nm trapping laser, where the top panels are snapshots from the videos recorded for directional cell movement (marked by red arrows) under the action of the laser beam (see Supplementary Movies), and the bottom panels depict the trap stiffness calculated using the standard optical tweezer tool [41]. Overall, since the RBCs in all osmotic conditions have a larger index of refraction than the background medium (n= 1.334)[22], they are pulled towards the beam center due to the gradient force. For suspended particles or cells with positive polarizability (such as RBCs), a larger refractive index contrast corresponds to a greater polarizability and consequently a larger gradient force. Therefore, the gradient force exerted on the RBCs follows a trend of hypertonic > isotonic > hypotonic, as directly measured from the optical tweezers (Figs. 3d-3f). By examining the motion of the trapped cells from the videos (Figs. 3a-3c), all of which were taken under identical trapping conditions, we found that the attraction due to the gradient force exerted on the RBCs clearly follows the same trend. Although the 532-nm laser beam (for the nonlinear propagation experiments of Fig. 1 and Fig. 2) can definitely apply the gradient force on the RBCs (see supplementary videos), it cannot create a stable trap for a single cell in our optical tweezer setting due to the strong scattering force at this wavelength. Thus, for a quantitative stiffness measurement, the infrared wavelength (960 nm) was used in the tweezer setup to directly characterize and compare the optical gradient forces that act on the RBCs under different conditions. To simplify our calculations, we assumed the RBCs as disk-shaped (prolate ellipsoid) objects in isotonic conditions and spherical objects in the hypotonic and hypertonic conditions with different average diameters (8.0 μm for isotonic, 9.6 μm for hypotonic, and 6.4 μm for hypertonic cells). Then, the trapping force can be estimated from the Langevin equation [45,46], where the corner frequency $f_{c,x} = \kappa_x/2\pi\gamma$ and trapping stiffness $\kappa_x$ can be evaluated (here, $\gamma = 6\pi\eta a$ is the particle friction coefficient, $\eta$ is the viscosity of the solution, and $a$ is the radius of the particle). The results obtained with single trapped cells are plotted in Figs. 3d-3f. Although the gradient forces can vary at 960-nm and 532-nm wavelengths, the force measurement results in Fig. 3 are only used as a guide for comparison, since the videos taken at both wavelengths clearly show that they exhibit identical trends. It is nontrivial to quantitatively measure and compare the forward-scattering force on the RBCs. We can infer that the force also follows the trend of hypertonic > isotonic > hypotonic from the measured power transmission and absorption spectrum under different osmotic conditions (see Supplementary Material). Interestingly, we found that with a low numerical aperture (NA=0.65) so that the gradient force is weaker (more comparable to the nonlinear propagation experiment), when the cells were attracted to the beam focus by the 532-nm laser, they were pushed out of the observation plane by the strong scattering force (see supplementary videos). This result is different from the case of the high NA setting, where multiple cells were quickly attracted by the gradient force to the focus to form clusters before they could be pushed away. In fact, as we show in our following theoretical analysis, the forward scattering force plays an essential role in the nonlinear self-guiding of light in biological suspensions. [21]

To better understand the physics of the optical force-mediated nonlinearity, we have developed a model to simulate the nonlinear beam propagation in bio-soft-matter (see Supplementary Material). Instead of a priori assumption of any particular form for the nonlinearity, we let the beam propagate in a dynamic waveguide, which forms due to the spatial variation of the particle concentration. The time evolution of the particle concentration distribution is modeled using a diffusion-advection equation, where the velocity field is determined by the intensity-dependent optical forces. Contrary to previous models, we consider that the particles are affected not only by an optical gradient force but also by a forward-scattering force [21,47] which pushes the particles along the beam propagation direction [48,49]. Without the scattering force, the mathematical description is reduced to the exponential nonlinearity model for the nanosuspension of particles in the steady-state limit [40,50]. However, the inclusion of a strong forward-scattering force causes a fundamentally different nonlinear response that is nonlocal in the propagation direction (where the particle concentration does not necessarily peak at the beam focus). To simulate the nonlinear self-focusing effects under different buffer conditions, we calculated the change in beam size for different gradient and scattering force parameters. The results are shown in Fig. 4, where Fig. 4a is a 3D plot of the beam size as a function of the forces, and Fig. 4b and Fig. 4c illustrate the decrease in beam size when the gradient or the scattering forces varied. The dashed white line in Fig. 4a corresponds to the minimal beam diameters achievable in simulations before beam collapse occurs. (In these simulations, we did not include the particle-particle interactions, thermal effects, and random scattering [40,50], which could suppress the collapse.) In the bottom panels, the beam dynamics through isotonic RBC-like particle suspensions with added random scattering is shown for both linear (low power of 10 mW) and nonlinear (high power of 350 mW) propagation based on parameters extracted from our experimental conditions. The change in size, volume and refractive index of the RBCs under different osmotic conditions accounts for the variations in magnitude of the optical forces, which modifies the optical nonlinearity [51]. These results are qualitatively consistent with the experimental observations.

**Discussion**

Finally, we discuss the reproducibility of our experimental results and the role played by free Hb in the optical nonlinearity. The results in Fig. 1 and Figs. 2(a, b) were readily reproduced in fresh blood samples, but the experiments using the same blood samples with identical concentrations after they had been stored in a refrigerator for a certain time period had different results. For example, Fig. 2c shows the measurements with the blood sample in Fig. 2b retaken after two weeks. (Initially, the fresh blood sample was centrifuged, and the supernatant was removed to ensure that only the RBCs would remain before different osmotic solutions were prepared.) Clearly, the beam dynamics vs. input power is very similar in Figs. 2b and 2c for isotonic and hypertonic solutions, but the difference is evident for the hypotonic condition: here, the beam self-traps to a significantly smaller size at a much lower input power (150 mW) than the fresh sample experiment in Fig. 2b (green curves). After counting the number of RBCs using a hemocytometer, we have found that the number of RBCs in the hypotonic buffer is now only one-third of the value in the isotonic and hypertonic buffers, which indicates that most of the hypotonic RBCs were lysed in the old sample and released free Hb. (The increase in membrane stiffness and hemolysis of RBCs with storage is known, and the cells more easily lyse in the hypotonic solution [38]). To prove that the enhanced nonlinear response from the aged hypotonic RBC suspension can be attributed to the released Hb, fresh RBCs were intentionally lysed in deionized water to prepare solutions at four RBC concentrations. The results are shown in Fig. 5a. The nonlinear self-trapping was clearly realized at a power of approximately 100 mW at all tested concentrations (2.4, 5.1, 8.6, and 15.0 x $10^6$ cells per mL), but a higher concentration caused stronger self-focusing (smaller minimum focused size). When the power wasfurther increased, the thermal effects dominated, so the beam dramatically expanded. A higher concentration also leads to a stronger thermal self-defocusing at 500 mW. Figures 5b-5e show the typical output transverse intensity patterns for the self-trapped beam and the formation of thermal defocusing rings at two different concentrations. The measured power transmission is approximately 50% (70%) for the low (high) concentrations, which is much higher than that in healthy RBC suspensions due to the reduced scattering loss. These results clearly demonstrate that free Hb can exhibit a nonlinear optical response, which results in the enhanced nonlinearity in aged hypotonic RBC suspensions. The underlying mechanism of the Hb-enhanced nonlinear response certainly merits further studies. Intuitively, we believe that it is similar to the optical force-induced nonlinearity in plasmonic nanosuspensions [54], since metallic ion nanoparticles are contained in the protein chain of hemoglobin. The measured response time of self-trapping in Hb solutions was approximately 200 ms at the laser power used. Thus, we do not think that it is driven by the thermophoresis effect like the "hot-particle" solitons in dielectric nanoparticle suspensions [44], since the latter typically requires much higher laser power and occurs at much slower time scales.

In conclusion, we have studied the nonlinear beam propagation in human RBCs suspended in isotonic, hypotonic, and hypertonic buffer solutions, and we have found that RBCs exhibit a strong self-focusing nonlinearity that can be controlled by the chemistry of the liquid buffer. In particular, the optical nonlinearity can be tuned via osmosis and increases with the osmotic pressure outside the cells in fresh blood samples. In aged blood samples, free hemoglobin from the lysed RBCs plays an active role in the optical nonlinearity and enhances the nonlinear response in hypotonic conditions. From direct video microscopy and optical tweezers measurement, we conclude that the trapping force that acts on the RBCs is the strongest in the hypertonic condition and the weakest in the hypotonic case. Building further on our prior work, we provide a theoretical model to explain the observed effects. Our demonstration of the controlled nonlinear self-trapping of light in RBC suspensions through their tunable optical nonlinearity can introduce new perspectives to the development of diagnostic tools, which is very promising towards future laser treatment therapies of blood-related diseases [52,53].

**Materials and Methods**

*Isolation and preparation of RBC samples*: The human blood samples in this study were obtained from anonymous donors through the Blood Centers of the Pacific, California. The blood samples were collected in ethylenediaminetetraacetic acid (EDTA) tubes and centrifuged at 3500 RPM for 5 minutes. Then, the supernatant was removed, and the remaining RBCs were washed 3 times in PBS buffer before the RBC suspensions were prepared for our experiment. To study the nonlinear optical response of RBCs suspended in different osmotic conditions, we intentionally let the RBCs disperse in the hypotonic and hypertonic buffers in addition to their natural isotonic condition. For direct comparison, we diluted 50 μL of RBCs in 4 mL volume of isotonic, hypotonic and hypertonic buffers to obtain a final concentration of approximately $10^5$-$10^6$ cells/mL as counted by a hemocytometer. As it is well known in the literature [22,28,39], "normal" RBCs exhibit a biconcave disc shape in the isotonic PBS buffer, "swollen" RBCs take a balloon-like sphere shape in the hypotonic buffer, and "shrunk" RBCs take an irregular spiky shape in the hypertonic buffer (as illustrated in the top row of Fig. 1). Different buffers were prepared by changing the NaCl concentration from 8.2 mg/mL (isotonic) to 4.0 mg/mL (hypotonic) and 14.0 mg/mL (hypertonic) while maintaining the concentration of three other salts unchanged (KCl: 0.201 mg/mL; $Na_2HPO_4$: 1.434 mg/mL; $KH_2PO_4$: 0.245 mg/mL). To ensure that there were little or no lysed RBCs that would release Hb into the solutions, the samples were centrifuged for 5 additional minutes at 3500 RPM immediately before the experiment.

*Nonlinear optical self-trapping experiment*: In the first set of experiments, a linearly polarized CW laser beam with a wavelength of 532 nm was collimated and subsequently focused (with a lens of 125-mm focal length) into a 3-cm-long glass cuvette filled with the RBC suspensions for different osmotic conditions as described above. In particular, the focused beam was approximately 28 μm in diameter at the focal point, which was located approximately 1 cm away from the input facet of the cuvette to avoid heating and surface effects[21]. As a consequence, the beam propagated by approximately 2 cm after the focal point through the suspension before exiting the output facet. Both linear and nonlinear outputs from the sample were monitored with a CCD camera and a power detector. To maintain unsaturated image detection, the necessary attenuation optical elements were used to adjust the illumination power before reaching the camera. Furthermore, the beam diameters were measured using the Beamview program after careful calibration. The experimental setup is similar to those in our previous work for the study of beam propagation in colloidal suspensions [21,40].

*Force measurement in the optical tweezers experiment*: In the second set of experiments, a home-built optical tweezers system [41] was used to measure the optical gradient force. In our tweezers setting, a CW laser beam ($\lambda$ = 960 nm) was coupled into the optical pathway of the microscope and subsequently tightly focused into the sample using an oil immersion objective lens (numerical aperture NA = 1.3). Separate samples that contained droplets of different RBC solutions were sandwiched between a microscope slide and a cover slip. The forward-scattered light from the trapped cells was collected by a condenser lens (NA = 1.2) and subsequently focused onto a position-sensitive detector (PSD). The PSD positional signals were acquired using a custom-made

LabVIEW program. The trap stiffness and gradient force were calculated from the corner frequency of the PSD using a standard optical tweezers tool [41]. To simplify the calculations, the hypotonic (and hypertonic) RBCs were treated as spherical objects, whereas isotonic RBCs were treated as disk-shaped (prolate ellipsoid) objects. In the latter case, a shape-dependent "particle" friction coefficient was used for the force calculation [55, 56]. In addition, we used a CCD camera to record the cell movements from different RBC solutions under a microscope with two objectives: NA =0.65 and NA= 1.3, as driven by the 960-nm laser beam (used in the tweezers experiment for the force calibration) or the 532-nm laser beam (used for the beam propagation experiment). These video results are presented in the Supplementary Material. In particular, they are not meant to show a stable single cell trapping but to illustrate the cell movement against Brownian motion under the action of optical forces that depend on the cell conditions (shape and size) and trapping beam.

## Acknowledgements

This work was supported by the NIH, NSF and ARO (USA), and by the National Key R&D Program of China (2017YFA0303800), the National Natural Science Foundation of China (91750204,11504184,11604058), , the NSERC through the Steacie, Strategic, Discovery and Acceleration Grants Schemes, and the Canada Research Chair Program (Canada). R.M. acknowledges additional support by the Government of the Russian Federation through the ITMO Fellowship and Professorship Program (grant 074-U 01) and from the 1000 Talents Sichuan Program in China.

## Conflict of interests

The authors declare no conflicts of interest. The authors declare no competing financial interests.

## Contributions

All authors contributed to this work.

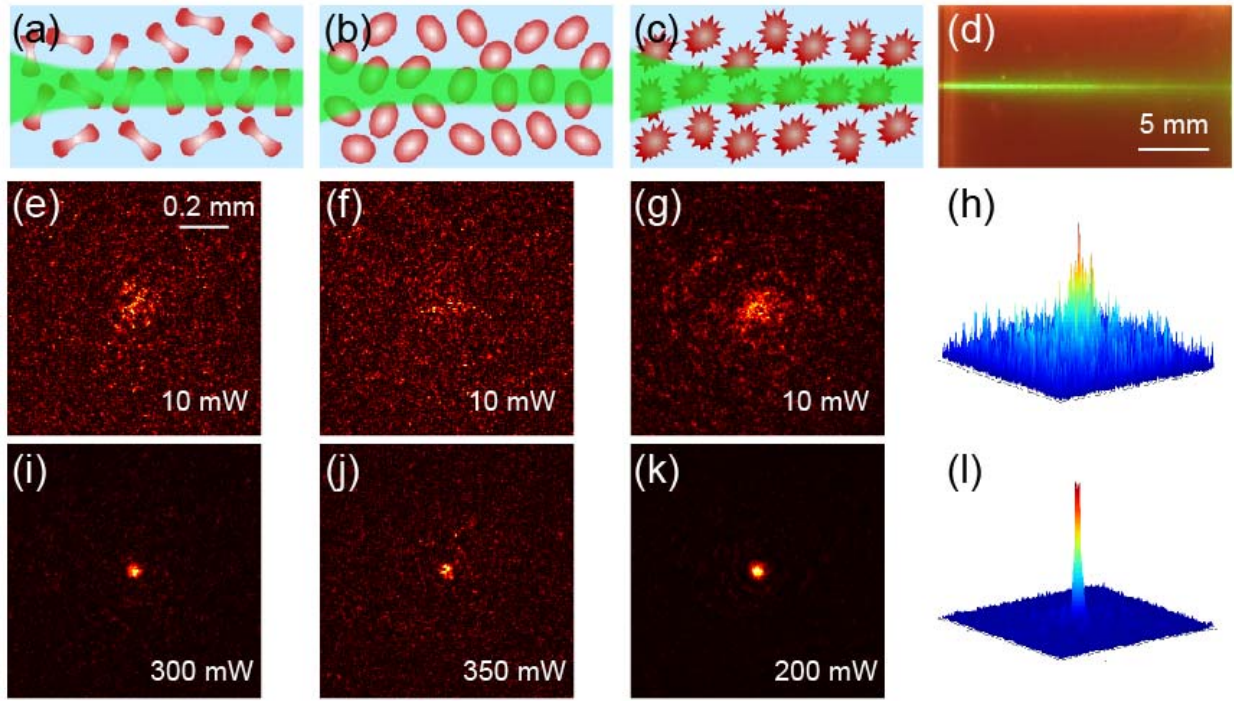

**Figure 1. Self-trapping of light through human RBC suspensions under different osmotic conditions.** (a-c) Illustrations of the beam dynamics in (a) isotonic, (b) hypotonic, and (c) hypertonic suspensions. (d) Side-view image of a self-trapped beam. (e-g) Observed output intensity patterns at a low power, which show the linear diffraction and strong scattering of the laser beam. (i-k) Corresponding patterns at a high power, which show the beam localization due to nonlinear self-trapping. (h, l) 3D plots of the intensity patterns corresponding to (g, k), respectively.

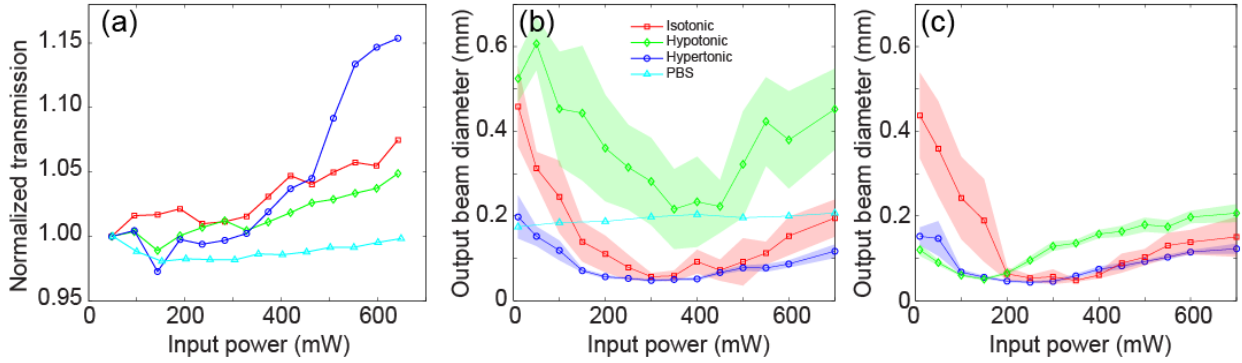

**Figure 2. Normalized transmission and output beam size as a function of input power.**
(a) Measurement of the normalized transmission and (b) output beam size change in fresh RBC suspensions of different buffer solutions. The cyan (triangle) curve depicts the results obtained from the PBS background solution without RBCs as a reference, which indicates no appreciable self-action of the beam in the buffer solution itself. The blue (circle), red (square), and green (diamond) curves show the data obtained from RBC suspensions in hypertonic, isotonic, and hypotonic solutions, respectively, where the error ranges in (b) are indicated by the shaded regions. (c) Corresponding results from the same blood sample but after the RBCs have been stored in a refrigerator for two weeks, where the nonlinear focusing is dramatically enhanced in the hypotonic solutions.

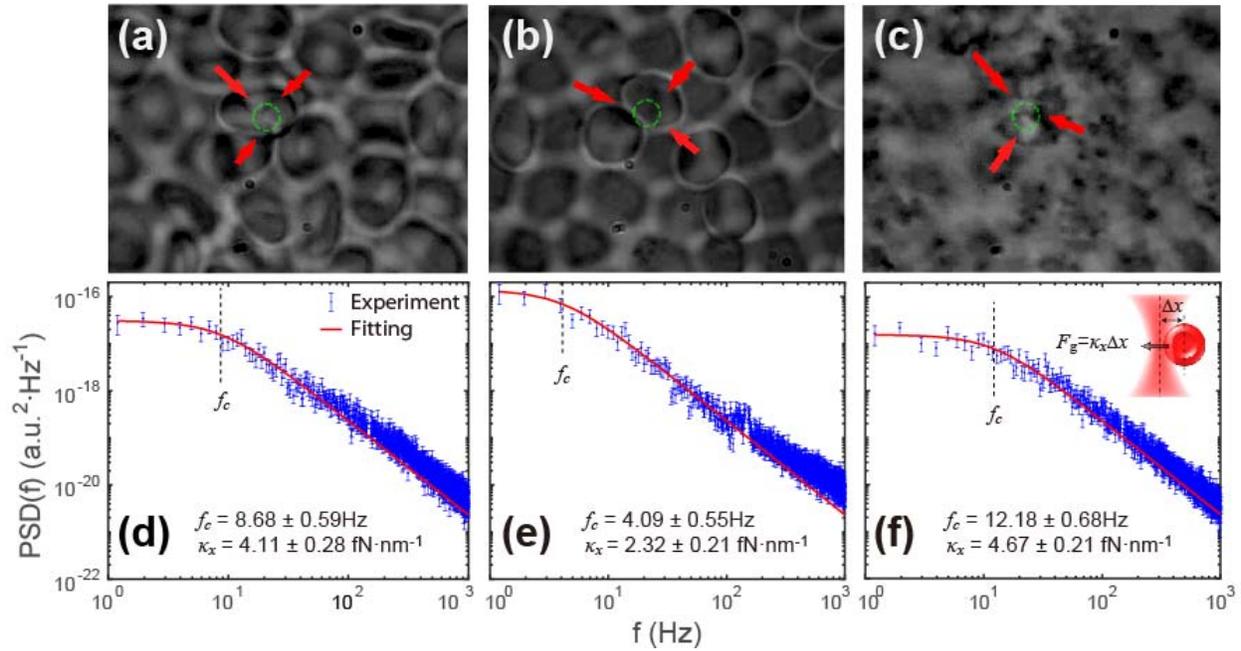

**Figure 3. Optical gradient forces on RBCs under different osmotic conditions examined by optical tweezers.** (a-c) Snapshots of RBC movement towards a 960-nm laser beam (position marked by a dashed green circle) in isotonic, hypotonic, and hypertonic solutions, respectively, as observed under a microscope. The red arrows illustrate the directional cell movement (see corresponding Media file in the Supplemental Material). (d-f) Power spectrum analyses showing the trap stiffness $\kappa_x$ of a single RBC from the three suspensions in accordance with (a-c), where the vertical dashed lines mark the corner frequency $f_c$. The inset in (f) illustrates a single RBC that moves into the trap under the action of gradient force.

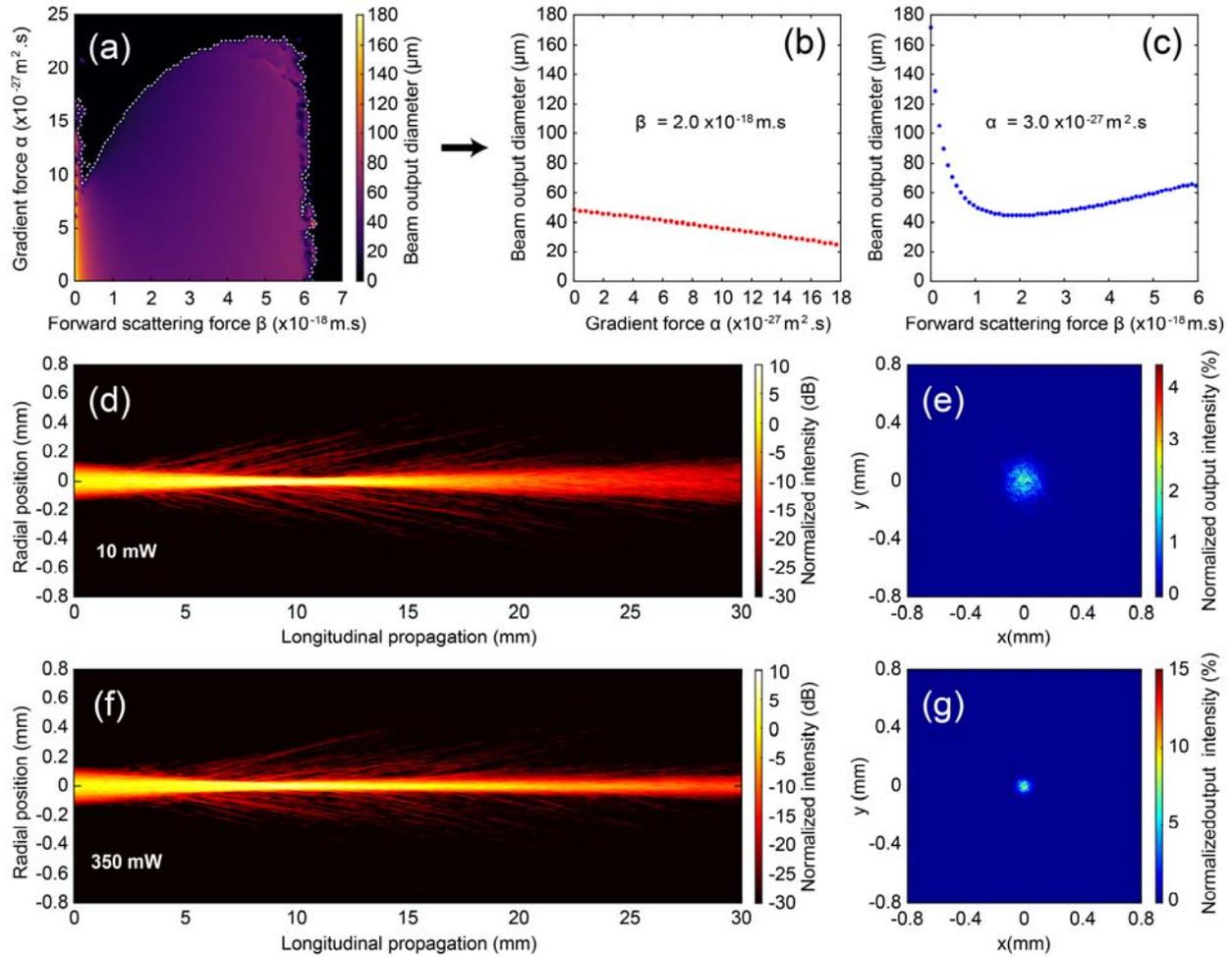

**Figure 4: Simulations of the optical force-induced nonlinear beam dynamics in RBC-like suspensions**. (a, b, c) Beam size (FWHM) change as a function of the gradient and scattering forces obtained via numerical simulations using a 350-mW input power and neglecting random scattering effects, where one observes the change in beam size when either the gradient or the scattering force is "turned off". (d, f) Side-view of the beam propagation and (e, g) corresponding output transverse intensity patterns after propagating through an RBC-like random scattering medium at low (d, e) and high (f, g) beam power. The beam side-views and output intensity patterns are normalized with respect to their respective maximal input powers.

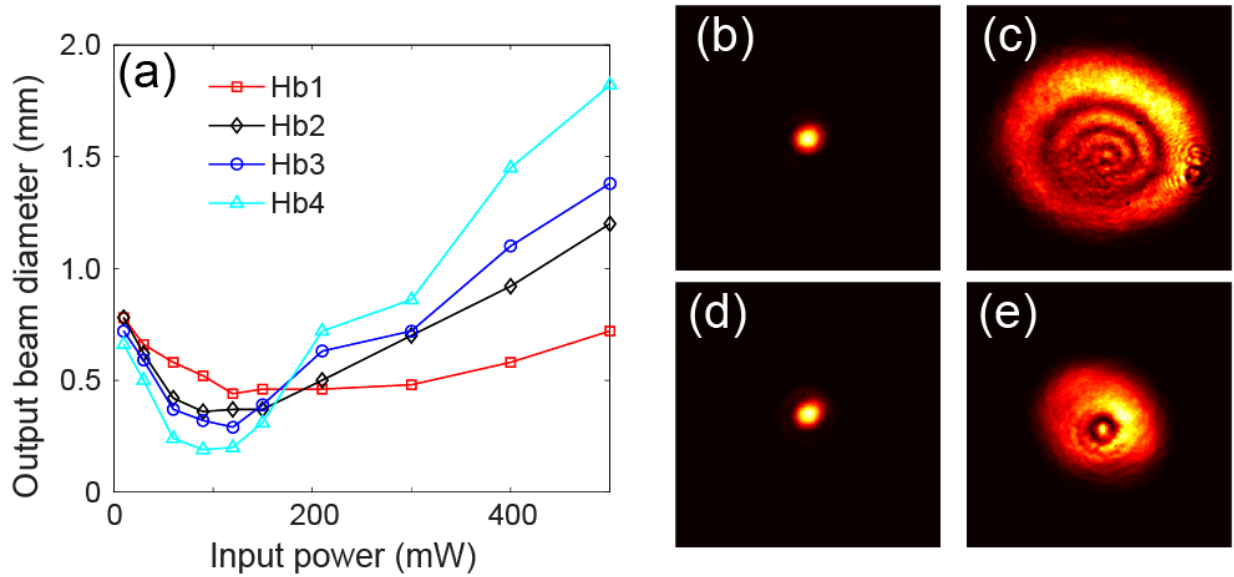

**Figure 5: Nonlinear optical response of lysed RBCs (free hemoglobin) in water.** (a) Output beam size as a function of input power through the Hb solutions for four different concentrations. The RBC concentrations for the four curves (Hb1-Hb4) are 2.4, 5.1, 8.6, and 15.0 million cells per mL. Nonlinear self-focusing of the beam occurs at approximately 100 mW for high concentrations of Hb, but it subsequently expands into thermal defocusing rings at high powers. (b-e) Typical output transverse intensity patterns taken for the self-trapped beam (b, d) and thermally expanded beam (c, e) for low (d, e) and high (b, c) concentrations.